\input{aipcheck}

\documentclass[
    ,final            
  ]
  {aipproc}

\layoutstyle{6x9}
\usepackage{amssymb}

\begin{document}

\title{Study of Planetary Systems Around Giant Stars}


\author{Matias I.\ Jones}{ 
  address={Departamento de Astronomia, Universidad de Chile; Camino El Observatorio 1515, Las Condes, Santiago, Chile} }

\author{James S.\ Jenkins}{ 
  address={Departamento de Astronomia, Universidad de Chile; Camino El Observatorio 1515, Las Condes, Santiago, Chile} } 

\author{Patricio Rojo}{ 
  address={Departamento de Astronomia, Universidad de Chile; Camino El Observatorio 1515, Las Condes, Santiago, Chile} }


\begin{abstract}
More than 450 exoplanets have currently been detected, most
of them by the radial velocity (RV) technique. While the majority of exoplanets have been found
around main-sequence (MS) FGK stars (M\,$\lesssim$ 1.5M$\odot$), only
a small fraction ($\sim$\,10$\%$) have been discovered orbiting post-MS stars.
However, such stars are known to host exoplanets and the detection fraction
appears to be larger than for solar-type dwarfs.  To date $\sim$ 30 planets have been
found orbiting giant stars which have revealed interesting properties that contrast with
the results found for solar-type stars.
We are carrying out a RV search for planets around giant stars in the southern hemisphere in order
to study different formation scenarios for planets around intermediate-mass stars
and the effect of the post-MS evolution of the host stars on the orbits of close-in planets ($a$ $\lesssim$ 0.6\,AU).
\end{abstract}

\maketitle


\section{Scientific Context}

To date more than 450 planetary candidates have been detected using the radial velocity (RV) technique. While most of the 
stars that host detected planets are main-sequence (MS) FGK stars (M $\lesssim$ 1.5 M$\odot$), only a small fraction ($\sim$ 10$\%$)
are sub-giants or giants.  However, evolved stars present an ideal case where to use the RV to search for planets around 
intermediate-mass stars (M $\gtrsim$ 1.5M$\odot$) because their are cooler and rotate slower (Schrijver $\&$ Plos 1993, A$\&$A, 278,
51) than their former MS progenitors and thus present more and narrower lines in their spectra. Also, post-MS stars with B-V < 1.2 
are quite stable and have a relatively low level of jitter (less than 20 m/s; Hekker et al.\ 2006, A$\&$A 454, 943). 
To date $\sim$50 planetary companions around post-MS stars with masses > 1.3 M$\odot$ have been detected and although the sample is 
limited, interesting results are emerging. \newline
Firstly, there is a lack of close-in orbits companions (with semi-major-axis $\lesssim$0.6 AU) around giant stars. Figure 1 shows the
semi-major-axis distribution for planets orbiting stars more massive than 1.3M$\odot$ in the MS stage (filled squares), the sub-giant 
phase (filled triangles) and the red giant branch (RGB; filled circles). As can be seen there is no planet orbiting giant stars
closer than $\sim$0.6 AU. This observational result might suggests that planets in close-in orbits are destroyed during the evolution
of the host stars through the RGB (see e.g.\ Villaver \& Livio 2009, ApJ, 705, 81). However, the semi-major-axis distribution of 
planets around subgiants suggests a different formation scenario around stars more massive than $\sim$1.5M$\odot$ rather than a host
star evolutionary effect, since the subgiants star have not expanded enough to destroy planets orbiting further than $\sim$0.1 AU 
(Johnson et al.\ 2007, ApJ, 665, 785). 
A second interesting result is that giant stars  harbouring planets do not show a trend  toward metal richness (see e.g.\ Doellinger et
al.\ 2009, A\&A, 505, 1311), in direct contrast to what  is observed in MS stars (Fischer \& Valenti 2005, ApJ, 622, 1102), which 
might suggests that the disk instability formation mechanism could be the dominant mechanism by which gas giant planets are formed 
in massive disks (Boss 2000, ApJ, 536, 101). However, this result could be also be explained as a pollution effect (Pasquini et al.\ 
2007, A\&A, 473, 979).
Finally, the fraction of giant stars harbouring giant planets is higher ($\sim$10$\%$) compared to solar-type stars ($\sim$5$\%$; 
Doellinger et al.\ 2009). This result suggests that giant planets are more efficiently formed in massive disks. 
Clearly a larger census of planets around post-MS stars is required to probe and understand these interesting emerging trends.

\begin{figure}
  \includegraphics[height=0.81\textwidth]{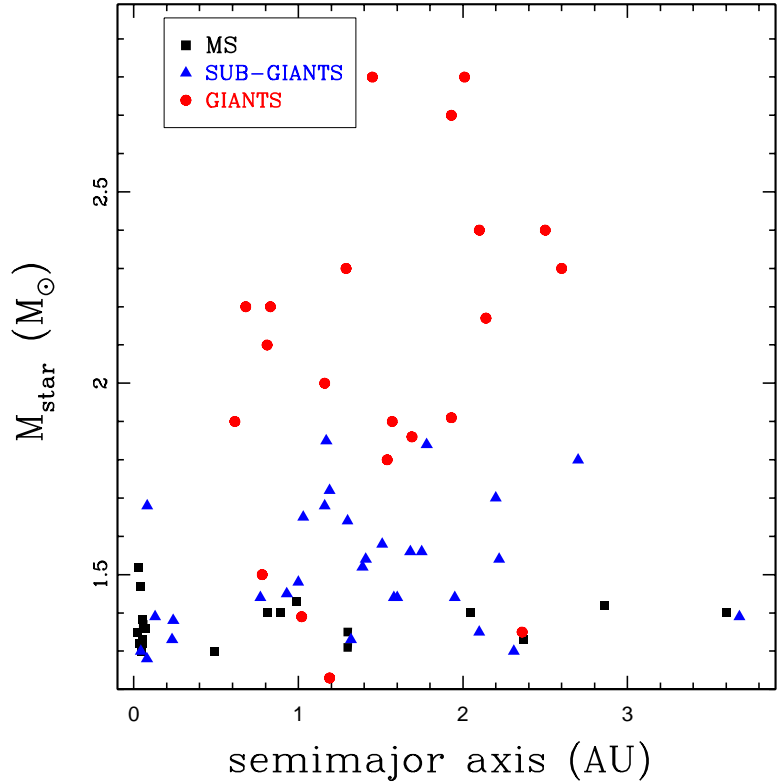}
  \caption{Semi-major-axis distribution for planets orbiting stars more massive than 1.3M$\odot$. 
  The squares, triangles and circles correspond to host stars in the MS, sub-giant phase and 
  giant branch, respectively.}
\end{figure}

\begin{figure}
  \includegraphics[height=0.8\textwidth]{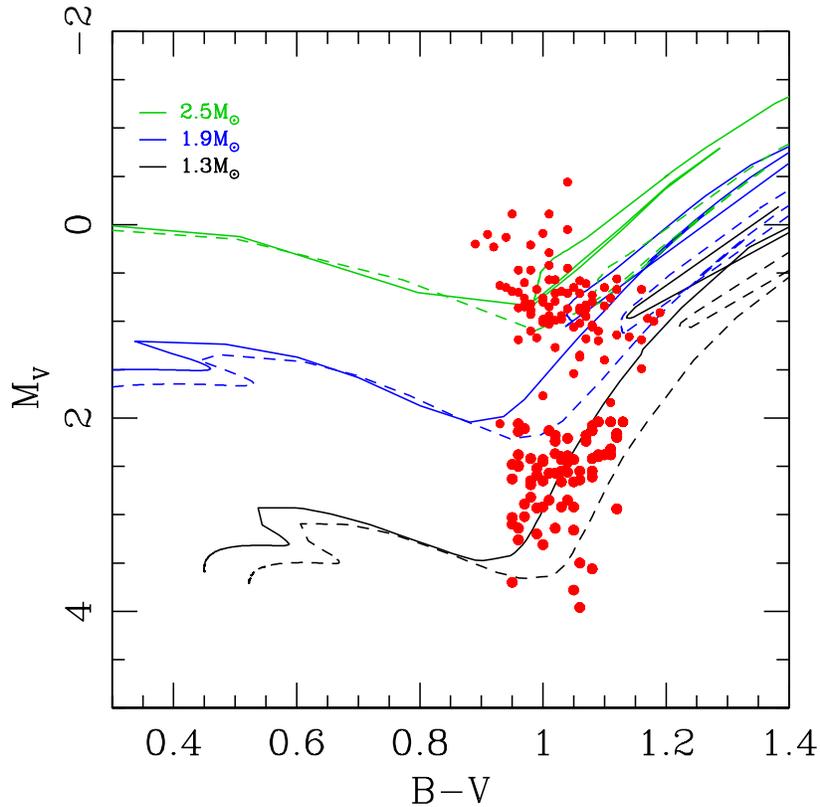}
  \caption{H-R diagram including our targets (filled circles). 
  Different evolutionary tracks (Marigo et al.\ 2008) are
  overplotted for stars with 1.3M$\odot$, 1.9M$\odot$ and 2.5M$\odot$ 
  (line pairs from top to bottom). 
  The solid lines correspond to evolutionary tracks with [Fe/H]=0.0 
  and the dashed lines to [Fe/H]=0.2. }
\end{figure}

\section{Targets}

We are monitoring a sample of ~150 bright giant stars (most of them brighter than V=7.0) in the Southern Hemisphere taken from the 
HIPPARCOS catalogue, according to their position in the HR diagram (RGB and Horizontal Branch (HB) stars) and precision in the 
parallax (better than $\sim$15\%). We also removed from the sample stars in binary systems and those with photometric variability 
greater than 0.015 mag. Figure 2 shows the position in the HR diagram for all of our targets (filled red circles). Also in Figure 2 
are overplotted evolutionary tracks from Marigo et al.\ 2008 (A\&A, 482, 883) for stars with ZAMS mass of 1.5M$\odot$ (black lines), 
1.9M$\odot$ (blue lines) and 2.5M$\odot$ (green lines). The solid lines correspond to evolutionary tracks with [Fe/H]=0.0 and the 
dashed lines with [Fe/H]=0.2.

\section{CURRENT STATUS OF THE PROJECT}

\subsection{Observations}
We are currently collecting data using the echelle spectrograph mounted on the 1.5m telescope at CTIO , which is equipped with an 
iodine (I2) cell that can be used to obtain precise wavelength calibrations (Butler et al.\ 1996, PASP, 108, 500) and therefore 
precise radial velocities. We are also using FEROS, mounted on the 2.2m telescope at La Silla Observatory. So far we have taken 
$\sim$10 spectra to each of our initial sample ($\sim$60 targets) and $\sim$2-6 spectra for the rest of the targets. 

\subsection{Data Reduction} 
We are developing a reduction pipeline  (in collaboration with D. Fischer; private communication) for the data taken at CTIO. We are
using quartz lamps taken with the I2 cell in the light path, in order to compute the instrumental profile (IP), which is necessary to
do the deconvolution of the I2 spectrum from the observed stars spectra. Finally, the radial velocities are computed using a 
cross-correlation between the observed spectra and a model template.

\subsection{Atmospheric Parameters} 
We have derived atmospheric parameters (effective temperature, surface gravity, microturbulence velocity and iron abundances) for 
each of the targets in our sample, which are used to determine their masses and ages by fitting evolutionary tracks. These results 
will be published soon in a paper. 

\begin{theacknowledgments}
Matias Jones acknowledges to Alma funding through grant 31080027 and Alma-Sochias for financial support. Patricio Rojo also acknowledges support from FONDECYT, Project 11080271.
\end{theacknowledgments}



\bibliographystyle{aipproc}   


\IfFileExists{\jobname.bbl}{}
 {\typeout{}
  \typeout{******************************************}
  \typeout{** Please run "bibtex \jobname" to optain}
  \typeout{** the bibliography and then re-run LaTeX}
  \typeout{** twice to fix the references!}
  \typeout{******************************************}
  \typeout{}
 }


\end{document}